# Accurate Interpolation of Ambient Noise Empirical Green's Functions by Denoising Diffusion Probabilistic Model and Implicit Neural Representation[†]

Guoyi Chen[1], Junlun Li*[1,2,3] and Bao Deng[1]

[1]State Key Laboratory of Precision Geodesy, School of Earth and Space Sciences, University of Science and Technology of China, Hefei 230026, China

[2]Mengcheng National Geophysical Observatory, University of Science and Technology of China, Mengcheng 233500, China

[3]Anhui Provincial Key Laboratory of Subsurface Exploration and Earthquake Hazard Risk Prevention, Hefei, 230031, China

*Corresponding author: lijunlun@ustc.edu.cn

---

# SUMMARY

Empirical Green's functions (EGFs) extracted from seismic ambient noise have been widely used to image Earth's interior structures, and the resolution of EGF-based tomography depends on the spatial density of seismic stations. However, due to cost and logistical constraints, it is often difficult to deploy dense seismic networks suitable for high-resolution tomography. While reliable interpolation of EGFs at unsampled locations could enhance tomographic resolution, the task remains inherently challenging and underexplored due to the dispersive nature of EGFs. In this study, we introduce DIER (diffusion-assisted implicit EGF representation), a self-supervised learning framework that integrates implicit neural representation with denoising diffusion probabilistic models to achieve high-fidelity EGF interpolation. In DIER, the diffusion process is conditioned on station coordinates to guide the transformation from random noise into EGF waveforms, which allows flexible reconstruction of five-dimensional EGF fields without labeled data or synthetic waveforms. We demonstrate the effectiveness of DIER through continent-scale EGF interpolation across the United States. The results show that DIER significantly outperforms the conventional radial basis function-based interpolation approach by generating EGFs with markedly improved phase alignment and dispersion characteristics. Surface wave tomography using the phase velocities derived from the interpolated EGFs also closely matches a reference model constructed from data acquired by a much denser seismic network. Our findings suggest that DIER provides a promising and cost-effective approach toward high-resolution ambient noise tomography in regions with sparse station coverage.

# 1. INTRODUCTION

Seismic ambient noise interferometry has become an indispensable tool for subsurface imaging in the last two decades (Wu et al., 2024). By cross-correlating continuous seismic noise recordings, empirical Green's functions (EGFs) primarily containing surface waves between station pairs can be effectively retrieved (Weaver, 2005). Then, using dispersive phase and group velocities in EGFs, it is possible to derive subsurface heterogeneities across different scales (Yao et al., 2006; Shen & Ritzwoller, 2016; Li et al., 2023), delineate fault geometries (Shapiro et al., 2005; Mordret et al., 2019), explore hydrocarbon or mineral resources (Mordret et al., 2013; Deng et al., 2022), monitor temporal velocity changes (Brenguier et al., 2008; Mao et al., 2025), among other geophysical applications. However, the quality and resolution of ambient noise tomography is highly influenced by density of seismic stations (Lin et al., 2013). Though advanced seismic sensing techniques (Zhan, 2020) have reduced instrumentational costs, deploying hundreds or even thousands of seismic stations remains economically and logistically prohibitive. Therefore, it is appealing to geophysicists to obtain virtual yet reliable EGFs at unsampled locations without actual field deployment.

As a well-established technique, seismic data interpolation is an essential step in standard processing workflows in exploration seismology. The strategies include prediction filters (Spitz, 1991), rank reduction (Oropeza & Sacchi, 2011), domain transformation (Abma & Kabir, 2006; Fomel & Liu, 2010), among others. Seismic data interpolation is also applied to teleseismic wavefields, which are typically characterized by sparse and irregular data sampling with low signal-to-noise ratios (SNR). For instance, by applying cubic-spline interpolation to resample three-component teleseismic recordings, Zhang & Zheng (2014) improved the structural delineation of the Moho at the Ordos block and adjacent regions in north China. Song et al. (2017) demonstrated that the radial basis function-based (RBF) interpolation has superior accuracy than cubic splines in receiver function interpolations. Zou et al. (2024) further validated the effectiveness of RBF-based interpolations in passive-source reverse time migration. In addition, curvelet transforms (Shang et al., 2017) and Delaunay tessellation (Yeeh et al., 2020) have also proven effective in reconstructing missing traces and improving quality of seismic imaging.

However, while data interpolation has been extensively applied to active-source exploration data and teleseismic body waves, few studies attempted to implement this technique for EGFs to our best knowledge. The primary reason lies in the inherent five-dimensional (time, the latitude and longitude coordinates of a station pair), irregularly-sampled, and dispersive nature of EGFs. Indeed, most of the aforementioned interpolation methods are specifically designed for regularly sampled 2D or 3D datasets containing non-dispersive body waves (Chen et al., 2019). Wang et al. (2025) introduced a 4-D reconstruction workflow that regularizes EGFs on gridded gathers to improve the stability of surface-wave dispersion measurements via rank-reduction filtering. However, this approach relies on discretizing station pairs on predefined bins, which may average out variability of recorded data within a binned space.

Recent advances in deep learning (DL) have achieved remarkable successes in a wide range of seismological applications, such as event phase picking (e.g., Zhu & Beroza, 2019; Mousavi et al., 2020; Chen & Li, 2022), seismic data denoising (e.g., Zhu et al., 2019; Trappolini et al., 2024), seismic inversion (e.g., Zhu et al., 2022; Liu et al., 2025), etc. Supervised and self-supervised DL approaches have also been explored for 2D and 3D seismic data interpolation, which show consistently higher reconstruction accuracy and fewer artifacts than conventional methods (Mousavi et al., 2024). Among the supervised methods, various network architectures have been explored to improve interpolation performance, such as generative adversarial network (Oliveira et al., 2018), autoencoder (Wang et al., 2020) and U-Net (Han et al., 2022). However, supervised learning often relies on high-quality synthetic datasets as labels to complement insufficient observed datasets. For EGFs, it is particularly challenging for numerical simulation to capture realistic characteristics of observed data due to dispersive nature of surface waves and uneven distribution of noise sources (Magrini & Boschi, 2020).

Meanwhile, self-supervised machine learning methods, which can infer missing records directly from observed data without labels, have attracted increasing attention. Representative approaches include implicit neural representation (INR) (Sitzmann et al., 2020), deep image prior (Kong et al., 2020), masked modeling (Yuan et al., 2022), and physics-informed neural networks (Brandolin et al., 2024). Among these approaches, INR shows great potentials for reconstructing high-dimensional data, and has been successfully applied to diverse fields, such as image processing (Sitzmann et al., 2020), view synthesis (Mildenhall et al., 2021), robotic perception (Chen et al., 2022), and seismic interpolation for both 2D teleseismic data (Gan et al., 2022) and 5D active-source seismic data (Liu et al., 2024; Gao et al., 2025). In INR, spatial or spatiotemporal coordinates are used as inputs, and the corresponding outputs of the neural network are inferred data at these specified coordinates. In particular, INR parameterizes the learning targets as continuous functions rather than on discretized grids, and thus is able to handle irregularly sampled data in arbitrary dimensions. Consequently, INR is well-suited for reconstructing 5D EGF wavefields sampled at irregular locations.

To better capture the complex dispersive nature of surface wave-dominated EGFs, deep generative models (Xu et al., 2015), which are a set of DL models dedicated to learning and reproducing complex data distributions, can also be incorporated into the interpolation framework of INR. Among different implementations of deep generative models, denoising diffusion probabilistic models (DDPMs, Ho et al., 2020) have emerged as a powerful tool to generate high-fidelity constrained data samples by progressively refining noise over multiple denoising steps. Liu et al. (2022), Ghosal et al. (2023) and Guan et al. (2024) demonstrated that DDPMs have remarkable generative performance in diverse data modalities. In 2D seismic data interpolation, Liu & Ma (2024) showed that DDPMs outperform both conventional methods and other sophisticated DL architectures in the supervised learning framework.

In this study, we propose diffusion-assisted implicit EGF representation (DIER), a novel deep learning framework that integrates INR with generative priors of diffusion models, to allow flexible and self-supervised high-fidelity interpolation of dispersive 5D EGFs. We first present the theory of DIER in detail, then evaluate its performance and benchmark against conventional RBF-based interpolations using ambient noise seismic data recorded in the continental United States (Hutko et al., 2017). Also, by comparing two surface-wave tomographic models (derived from dispersion curves from interpolated and original EGFs from a sparse array) with a reference model (derived from dispersion curves from a denser array), the viability of using DIER to improve resolution and fidelity of seismic tomography is also demonstrated.

## 2. METHODOLOGY

### 2.1 Probabilistic implicit neural representation for EGF wavefield

EGFs are retrieved by cross-correlating long-term seismic ambient noise between a pair of stations. The noise cross-correlation function $\Psi$ can be expressed as:

$$\Psi(\phi_i, \lambda_i, \phi_j, \lambda_j, \tau) = \int u(\phi_i, \lambda_i, t) u(\phi_j, \lambda_j, t + \tau) dt, \qquad (1)$$

where $u(\phi_i, \lambda_i, t)$ and $u(\phi_j, \lambda_j, t)$ are the continuous ambient noise records at stations $i$ and $j$, respectively, $(\phi_i, \lambda_i)$ and $(\phi_j, \lambda_j)$ are the latitude and longitude coordinates of respective stations, and $\tau$ is the time lag. The empirical Green's function $\hat{G}$ is then obtained by taking the derivative of $\Psi$ with respect to time (Shapiro & Campillo, 2004):

$$\hat{G}(\mathcal{C}, \tau) = \frac{d\Psi}{d\tau} = -\hat{G}_{ij}(\mathcal{C}, \tau) + \hat{G}_{ji}(\mathcal{C}, -\tau), \qquad (2)$$

where $\mathcal{C} = \{\phi_i, \lambda_i, \phi_j, \lambda_j\}$ represents the station coordinates or the input condition in a neural network, $\hat{G}_{ij}(\mathcal{C}, \tau)$ and $\hat{G}_{ji}(\mathcal{C}, -\tau)$ denote the causal and anti-causal parts of the EGFs.

The observed EGFs can be regarded as sparse samples of a continuous wavefield $\hat{G}$ defined in a five-dimensional space $(\phi_i, \lambda_i, \phi_j, \lambda_j, \tau)$. INR can approximate the EGF wavefield using an end-to-end mapping with neural network $f_\theta$, which deterministically learn the continuous mapping from coordinates to data values:

$$\hat{G}(\mathcal{C}, \tau) \approx f_\theta(v), \qquad (3)$$

where $\theta$ denotes learnable parameters of the network, and $v$ represents arbitrary coordinates as the input of $f_\theta$. Network $f_\theta$ can be constructed with any appropriate network architectures, such as multilayer perceptrons (Mildenhall et al., 2021) or convolutional neural networks (Chen et al., 2021). Early INR implementations typically use a full coordinate vector to predict a single data point (e.g. Tancik et al., 2020). However, due to the significant training cost and noise sensitivity of the pointwise INR, recent studies attempting to predict data blocks from partial coordinates have reported improved performance (Chen et al. 2021; Gao et al., 2025). In this study, since we focus on capturing the spatial continuity of EGF wavefields, we set $v = \mathcal{C}$ and predict the EGF waveform for a specified station pair as the network output, rather than using the full spatiotemporal input $(\mathcal{C}, \tau)$. Within the conventional deterministic INR framework, $f_\theta$ is optimized by minimizing the following loss function $L_d$:

$$L_d = \sum_{n=1}^{N} E\left(f_\theta(\mathcal{C}_n), \hat{G}(\mathcal{C}_n, \cdot)\right), \qquad (4)$$

where $E$ is a distance metric measuring the mismatch between real EGFs and predictions of the network in $L_1$ or $L_2$ norm, $\mathcal{C}_n \in \mathbb{R}^4$ $(n = 1, \ldots, N)$ represent the station coordinates for $N$ observational station pairs, and $\hat{G}(\mathcal{C}_n, \cdot)$ is the recorded EGF waveform for station pair $\mathcal{C}_n$.

Using the state-of-the-art DDPM as the backbone to learn the conditional distribution of EGFs $q(\hat{G}|\mathcal{C})$, we further reformulate coordinate-based mapping as conditional probabilistic modeling as:

$$q(\hat{G}|\mathcal{C}) \approx p_\theta(\hat{G}|\mathcal{C}), \qquad (5)$$

where $p_\theta$ is a parameterized conditional model, i.e., DDPM in this study. DDPM is adopted primarily because it can reliably model high-dimensional, complex data distributions, and allows DIER to accurately capture the dispersive characteristics of EGFs. Instead of minimizing $L_d$ directly (Eq. 4), we estimate $\theta$ by maximizing the log-likelihood of $p_\theta(\hat{G}|\mathcal{C})$:

$$\underset{\theta}{\mathrm{argmax}}\ \log p_\theta(\hat{G}|\mathcal{C}). \qquad (6)$$

An EGF for an arbitrary station pair $\mathcal{C}$ can be interpolated from the learned posterior distribution $p_\theta(\hat{G}|\mathcal{C})$. Details of DDPM as the backbone of DIER is provided in the following section.

## 2.2 Diffusion Denoising Probabilistic models

DDPMs approximate arbitrary data distributions via a two-stage Markov process. First, a forward Markov process is adopted to gradually contaminate data with Gaussian noises over $T$ steps:

$$q(\hat{G}^{1:T}|\hat{G}^0) \approx \prod_{t=1}^{T} q(\hat{G}^t|\hat{G}^{t-1}), \qquad (7)$$

where $\hat{G}^0$ is sampled from the distribution of observed EGFs in the training dataset, $\hat{G}^t$ (for $t = 1,2,\dots,T$) represents the EGF after the $t^{th}$ noising step, the total diffusion step $T$ is set to 500 in this study, $q(\hat{G}^t|\hat{G}^{t-1})$ describes the probabilistic transition of EGFs from step $t-1$ to step $t$ during the forward diffusion process, which is defined as:

$$q(\hat{G}^t|\hat{G}^{t-1}) = \mathcal{N}(\hat{G}^t; \sqrt{1-\beta_t}\hat{G}^{t-1}, \beta_t I), \quad (8)$$

where $\mathcal{N}$ represents a Gaussian distribution, $I$ denotes the identity matrix, $\beta_t$ ($t = 1, \ldots, T$) are a set of predefined noise schedule which typically increases from 0.0001 to 0.02 linearly (Ho *et al.*, 2020). By recursively applying Eq. (8), the forward process can be expressed as:

$$q(\hat{G}^t|\hat{G}^0) = \mathcal{N}(\hat{G}^t; \sqrt{\bar{\alpha}_t}\hat{G}^0, (1-\bar{\alpha}_t)I), \quad (9)$$

where $\alpha_t = 1 - \beta_t$ and $\bar{\alpha}_t = \prod_{s=1}^{t} \alpha_s$. Eq. (9) indicates that $\hat{G}^t$ can be sampled directly at any step by injecting scaled Gaussian noise in the original data. Additionally, as $t$ approaches $T$, $\bar{\alpha}_t$ approaches 0, and data are transformed into pure Gaussian noise, i.e., $\hat{G}^T \sim \mathcal{N}(0, I)$.

The reverse process is achieved by constructing a learnable probabilistic transition $p_\theta$ to iteratively denoise and reconstruct $\hat{G}^0$:

$$p_\theta(\hat{G}^{t-1}|\hat{G}^t) = \mathcal{N}(\hat{G}^{t-1}; \mu_\theta(\hat{G}^t, t), \sigma_t^2 I), \quad (10)$$

where $\mu_\theta$ and $\sigma_t$ represent the learnable mean and time-dependent variance of the Gaussian distribution for the reverse diffusion step $p_\theta$. Following Ho *et al.* (2020), we set $\sigma_t^2 = \beta_t$ and define a network $\varepsilon_\theta$ to reparametrize $\mu_\theta$:

$$\mu_\theta(\hat{G}^t, t) = \frac{1}{\sqrt{\alpha_t}}\left(\hat{G}^t - \frac{\beta_t}{\sqrt{1-\bar{\alpha}_t}}\varepsilon_\theta(\hat{G}^t, t)\right), \quad (11)$$

where $\varepsilon_\theta$ estimates the noise level in $\hat{G}^t$ and is optimized through the loss function $L$ by maximizing the log-likelihood of $p_\theta(\hat{G}^0)$:

$$L(\theta) = \mathbb{E}_{\hat{G}^0 \sim p_{data}, \varepsilon \sim \mathcal{N}(0,1)} \left[\left\|\varepsilon - \varepsilon_\theta(\sqrt{\bar{\alpha}_t}\hat{G}^0 + \sqrt{1-\bar{\alpha}_t}\varepsilon, t)\right\|^2\right], \quad (12)$$

where $p_{data}$ denotes the distribution of observed EGFs in the training dataset, $\varepsilon$ is the random noise sampled from the standard Gaussian distribution. Noise at each timestep can be predicted by the optimized $\varepsilon_\theta$ to retrieve $p_\theta(\hat{G}^0)$ progressively.

In DIER, the target is to generate EGFs conditioned on station coordinates $\mathcal{C} = \{\phi_i, \lambda_i, \phi_j, \lambda_j\}$. To obtain the conditional distribution $p_\theta(\hat{G}^0|\mathcal{C})$, we adopt classifier-free guidance (Ho & Salimans, 2022) by introducing a modified noise predictor $\hat{\varepsilon}_\theta$:

$$\hat{\varepsilon}_\theta = (1+k)\varepsilon_\theta(\hat{G}^t, \mathcal{C}, t) + k\varepsilon_\theta(\hat{G}^t, \mathcal{C}^0, t), \tag{13}$$

where $k$ is the guidance scale and set to 1 via trial and error, $\mathcal{C}^0$ denotes the null condition (all coordinates replaced with zeros). To support inference under the null condition $\mathcal{C}^0$, the conditioning information $\mathcal{C}$ is randomly dropped with probability $p_{drop}$ during training. The loss function $L$ thus is adjusted to:

$$L(\theta) = \mathbb{E}_{\hat{G}^0 \sim p_{data}, \varepsilon \sim \mathcal{N}(0,1)} \left[ \left\| \varepsilon - \varepsilon_\theta(\sqrt{\bar{\alpha}_t}\hat{G}^0 + \sqrt{1-\bar{\alpha}_t}\varepsilon, \tilde{\mathcal{C}}, t) \right\|^2 \right], \tag{14}$$

where $\tilde{\mathcal{C}} = \mathcal{C}$ with probability $(1 - p_{drop})$, and $\tilde{\mathcal{C}} = \mathcal{C}^0$ with probability $p_{drop}$. Following Rombach *et al.* (2022), we set $p_{drop} = 0.1$. The noise predictor $\varepsilon_\theta$ is optimized with a batch size of 256 for one million training steps. The Adam optimizer is used in the optimization (Kingma & Ba, 2014), which starts with a learning rate of 1e-4 and linearly decays to 0 with iterations.

Fig. 1 illustrates the framework of DIER. A 1D U-Net (Ronneberger *et al.*, 2015) is used as $\varepsilon_\theta$ to predict the noise at each timestep (blue arrows). The station coordinates $\mathcal{C}$ and timestep $t$ are jointly embedded into several up-sampled feature maps $\gamma$ of the U-Net via MLP layers:

$$\gamma' = f_{\theta_\mathcal{C}}(\mathcal{C}) * \gamma + f_{\theta_t}(t), \tag{15}$$

where $f_{\theta_\mathcal{C}}$ and $f_{\theta_t}$ denotes the MLP network with parameters $\theta_\mathcal{C}$ and $\theta_t$, respectively, and $\gamma'$ denotes augmented features embedded with information of coordinates and timesteps. This joint embedding allows the U-Net to receive spatially informed guidance at every reverse diffusion step and to progressively remove noise while preserving the station geometry. Previous studies have demonstrated that different coordinate embedding strategies can lead to diverse reconstruction fidelity (e.g., Tancik et al., 2020; Müller et al., 2022). After evaluating three candidate embedding strategies, we find that direct injection of absolute latitudes and longitudes into the MLP leads to the most generalizable performance (i.e., $f_{\theta_C}(\mathcal{C}) = f_{\theta_C}(\phi_i, \lambda_i, \phi_j, \lambda_j)$). Detailed comparison of different coordinate embedding strategies is provided in Section 5.1.

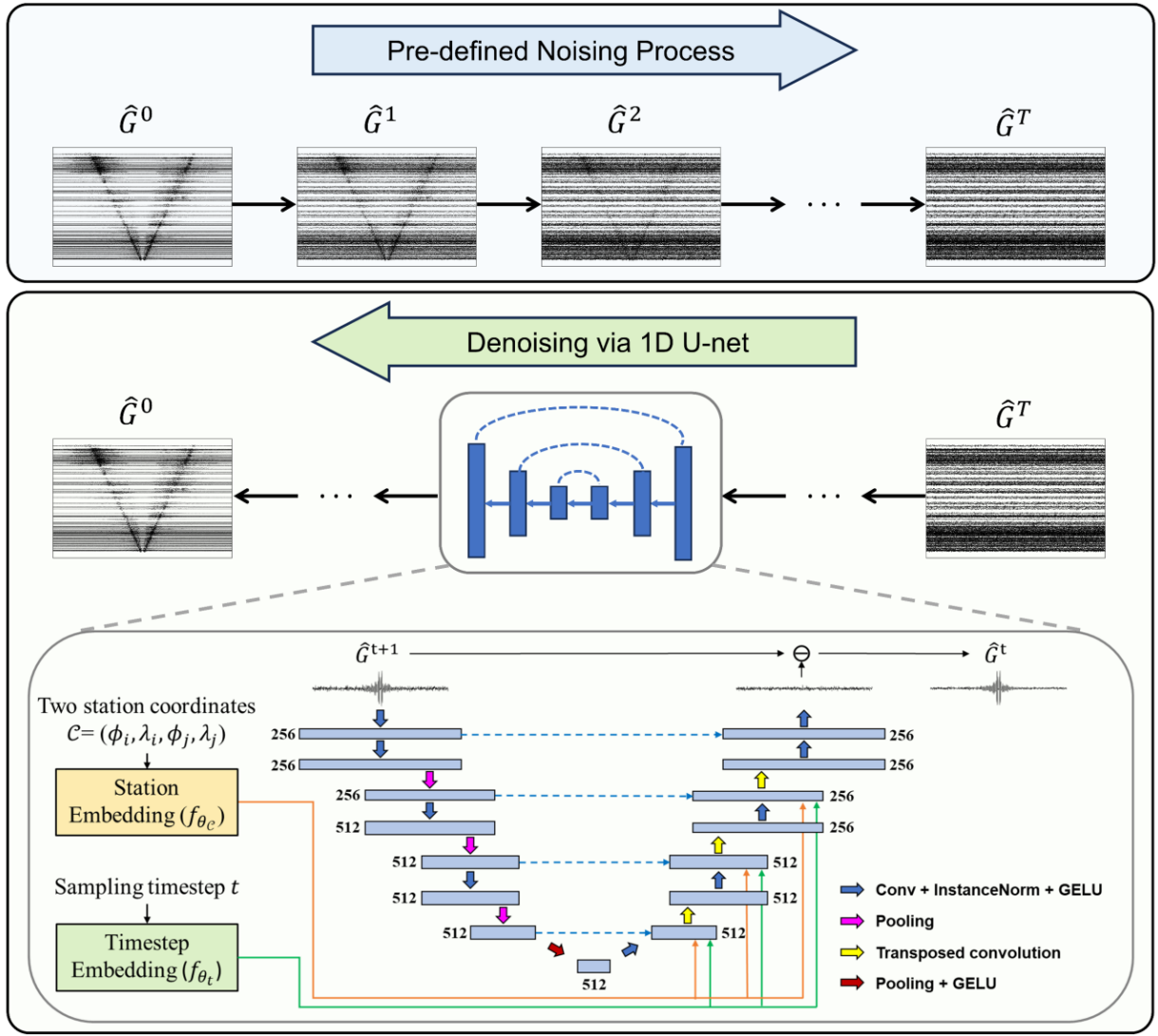

**Figure 1.** Schematic diagram of DIER. The blue and green bold arrows show the forward and reverse processes of DDPM, respectively. In the 1D U-Net, embedded station coordinates (orange arrows) and sampling timestep (green arrows) are injected to guide the denoising process. The blue dashed arrows indicate skip connection in the U-Net architecture, short arrows with different colors denote different network operations. Blue rectangles represent feature maps, with the number of channels indicated alongside.

## 3. DATASET AND PREPROCESSING

In this study, we use EGFs from the global empirical Green's tensor database curated by EarthScope Data Services (Hurko et al., 2017), which provides pre-calculated ambient noise EGFs from continuous seismic recordings at continental and global scales. We select the continental scale EGF dataset for North America, which is derived primarily from stations selected from the USArray, supplemented by regional and temporary broadband stations distributed across North America. We restrict our selection to seismic stations located within the continental U.S., yielding an EGF dataset from 460 stations in total (Fig. 2). The stations are divided into two groups: 400 randomly selected stations (blue dots in Fig. 2) serve as the training stations and the rest 60 stations (red dots in Fig. 2) are designated as the test stations. The EGFs from the training stations constitute the training dataset; the test dataset comprises both EGFs between two test stations and EGFs between one test and one training station.

The frequency of the raw EGFs ranges from 8 s to 300 s, and the record length is 7,200 s for the casual and anti-causal parts combined at a sampling rate of 1 Hz. To obtain signals with high SNR for method validation, we apply a bandpass filter to suppress high-frequency noise and low-frequency component with weak spectral energy, thereby enhancing waveforms in the 20-50 s frequency range. The EGF waveforms are subsequently downsampled to 0.25 Hz, truncated to the first 2,000 s from zero time, and multiplied by temporal Gaussian windows to emphasize surface waves (Fig. 3). The Gaussian windows are defined as $[d/v_{max}, d/v_{min}]$, where $d$ is the interstation distance, $v_{min} = 2$ km s$^{-1}$ and $v_{max} = 6$ km s$^{-1}$ are the presumed minimum and maximum phase velocities. Only EGFs with SNR>1.1 (defined as the ratio of the maximum absolute amplitude within the Gaussian window to that outside it) are retained for subsequent training and analysis. Each EGF waveform is then normalized by its maximum absolute amplitude to stabilize training of DIER. In total, the training and test datasets contain 35,867 and 11,785 EGFs, respectively. Notably, since surface wave tomography primarily relies on phase information rather than absolute amplitudes, this normalization does not affect the resulting tomographic models.

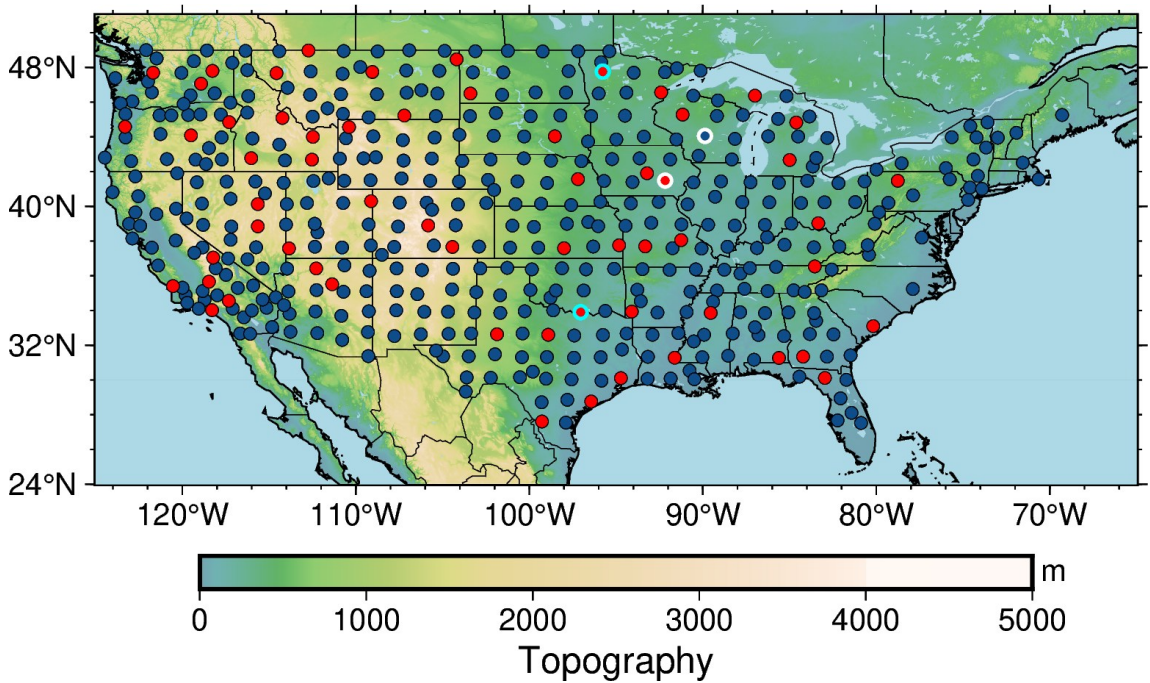

**Figure 2.** Distribution of seismic stations used in this study. EGFs between two blue stations are used for DIER training, and EGFs between two red stations, or between a red and a blue station are used for network testing. White and cyan bold circles highlight the station pairs related to the EGFs shown in Figs 5 and S1, respectively.

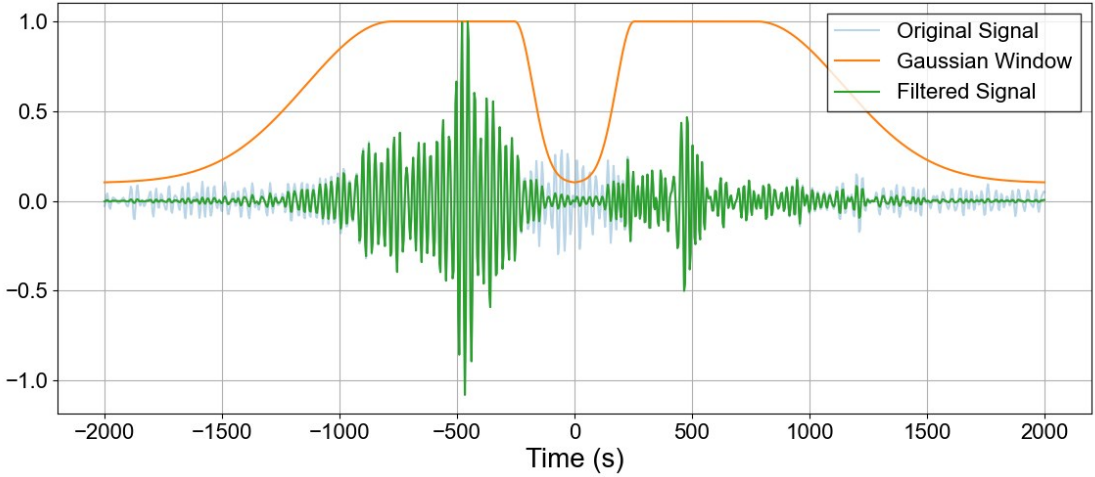

**Figure 3.** Example of Gaussian windowing applied to an EGF. The orange curve represents the temporal Gaussian window, which preserves signals within the expected surface-wave arrival window; the light blue trace shows the original waveform before filtering, and the green trace is the filtered waveform.

## 4. RESULTS

### 4.1 Performance of DIER Compared to Conventional RBF-based Interpolation

The RBF-based interpolation represents the state-of-the-art approach for reconstructing unmeshed or irregularly sampled data, and has been widely used in seismological studies (e.g., Song et al., 2017; Zou et al., 2024). In this section, we compare the performance of DIER with RBF-based interpolation for EGFs. We follow the same parametric configuration for RBF-based interpolation as used by Zou et al. (2024), the effectiveness of which has been validated in teleseismic body wave reconstruction. A brief overview for RBF-based interpolation formulations is provided in Appendix A. Fig. 4 shows an example of EGFs interpolated by RBF and DIER at randomly selected test stations. While RBF-based interpolation can reproduce the general characteristics of waveform envelopes with varied interstation distances (Fig. 4a), the conventional approach fails to capture subtle variations in surface wave phases and yields incorrect phase velocities. In comparison, DIER successfully reconstructs accurate dispersive characteristics for varied interstation distances (Fig. 4b).

Figs 5 and S1 show the comparison of the dispersion characteristics of the real (observed) and reconstructed EGFs. The locations of the corresponding station pairs are highlighted in Fig. 2. Fig. 5(a) shows the phase and group velocity dispersion spectrograms of a real EGF, and Fig. 5(b) shows the comparison between the real EGF and EGF interpolated with RBF. The distinct mismatch suggests that RBF-based interpolation fails to capture the frequency-dependent dispersive behavior intrinsic to surface wave propagation. The phase and group velocity spectrograms shown Fig. 5(c) also evidently demonstrate the interpolated EGF fail to capture the critical dispersive characteristics. In comparison, the EGF interpolated with DIER reproduces almost identical dispersive characteristics of the real EGF (Figs 5d and e). A similar conclusion can be drawn from another example shown in Fig. S1, where the quality of waveforms interpolated by RBF or DIER is inferior to those in Fig. 5. Nevertheless, even in the more challenging case, DIER still produces an EGF that align more closely with the real EGF.

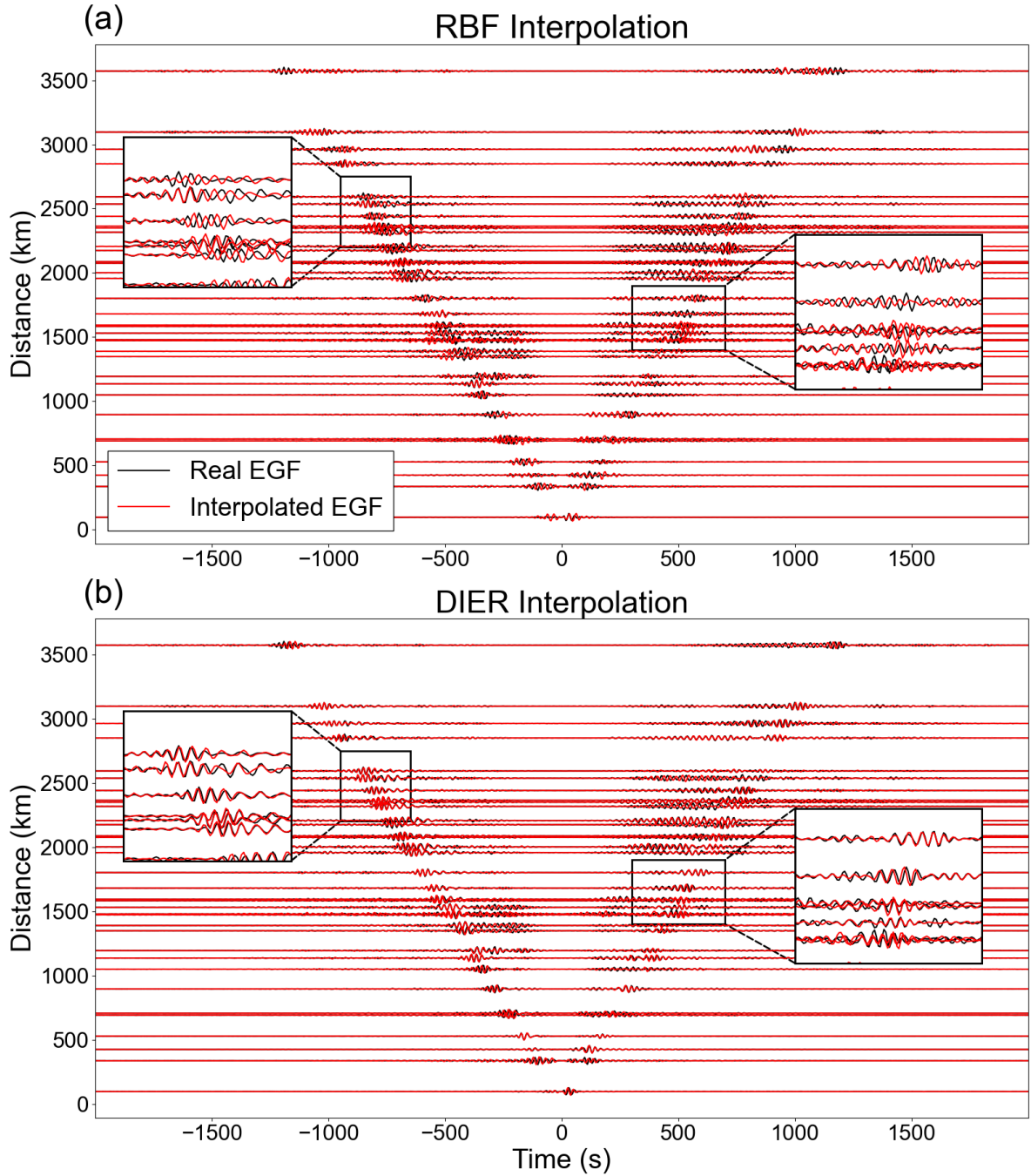

**Figure 4.** Comparison of EGF interpolation results. (a) Comparison between real EGFs and EGFs interpolated with RBF; (b) Comparison between real EGFs and EGFs interpolated with DIER. In each subfigure, the black lines indicate the real EGFs, and the red lines indicate the interpolation results. Two zoomed-in panels highlight waveform details within selected windows.

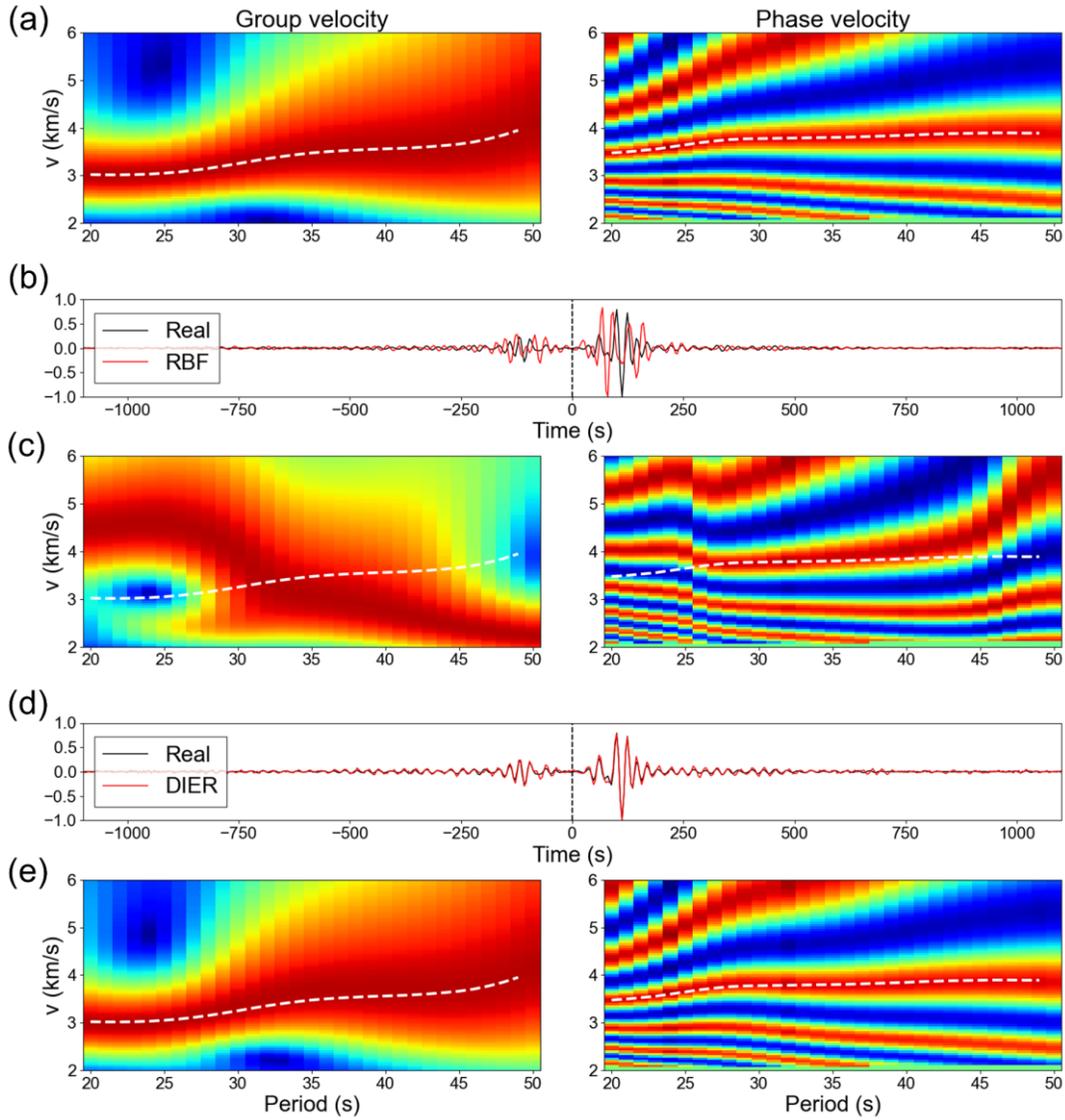

**Figure 5.** Comparison of dispersive characteristics of real and interpolated EGFs. (a) Group (left panel) and phase velocity (right panel) spectrograms calculated from the real EGF; (b) comparison between the real EGF (black line) and EGF interpolated by RBF (red line); (c) group (left) and phase (right) velocity spectrograms calculated from the EGF interpolated by RBF; (d) comparison between the real EGF and the EGF interpolated by DIER; (e) group and phase velocity spectrograms calculated from the EGF interpolated by DIER. The picked group and phase velocity

dispersion curves (white dashed lines) for the real EGF in (a) are replicated in (c) and (e) for comparison.

To further quantify the interpolation accuracy, we compare the statistical metrics of the EGFs interpolated by DIER and RBF at varying interstation distances (Fig. 6). The metrics in Fig. 6 are computed from all EGFs in the test dataset, with the casual and anti-casual parts evaluated separately due to their frequent asymmetry. Three indicators are evaluated, including the zero-lag cross-correlation coefficient between real and interpolated EGFs, the maximum cross-correlation coefficient and the corresponding time shift. These metrics measure raw waveform similarity, overall similarity after compensating for small arrival delays, and the temporal bias between the interpolated and real EGFs, respectively. The detailed definitions of the metrics are provided in Appendix B. As shown in Fig. 6(a), the zero-lag cross-correlation coefficients of EGFs interpolated with DIER are consistently higher, and those interpolated with RBF are close to zero or even negative for many station pairs, suggesting that the RBF-based interpolation substantially distorts the phase information. The comparison of the maximum cross-correlation coefficients with time shift, however, indicates that the discrepancy between the interpolated results by DIER and RBF appears smaller (Fig. 6b). The improved similarity after time compensation suggests that RBF-based interpolation reproduces the general waveform characteristics, but fails to recover correct phase velocities. As shown in Fig. 6(c), the time shifts required to align the RBF-interpolated EGFs with the real ones are significantly larger than those for DIER-interpolated EGFs, which are close to zero. Fig. 6(d) shows the counts of EGFs evaluated at each distance range. Overall, EGFs interpolated with DIER are more accurate than those interpolated with RBF for all distances.

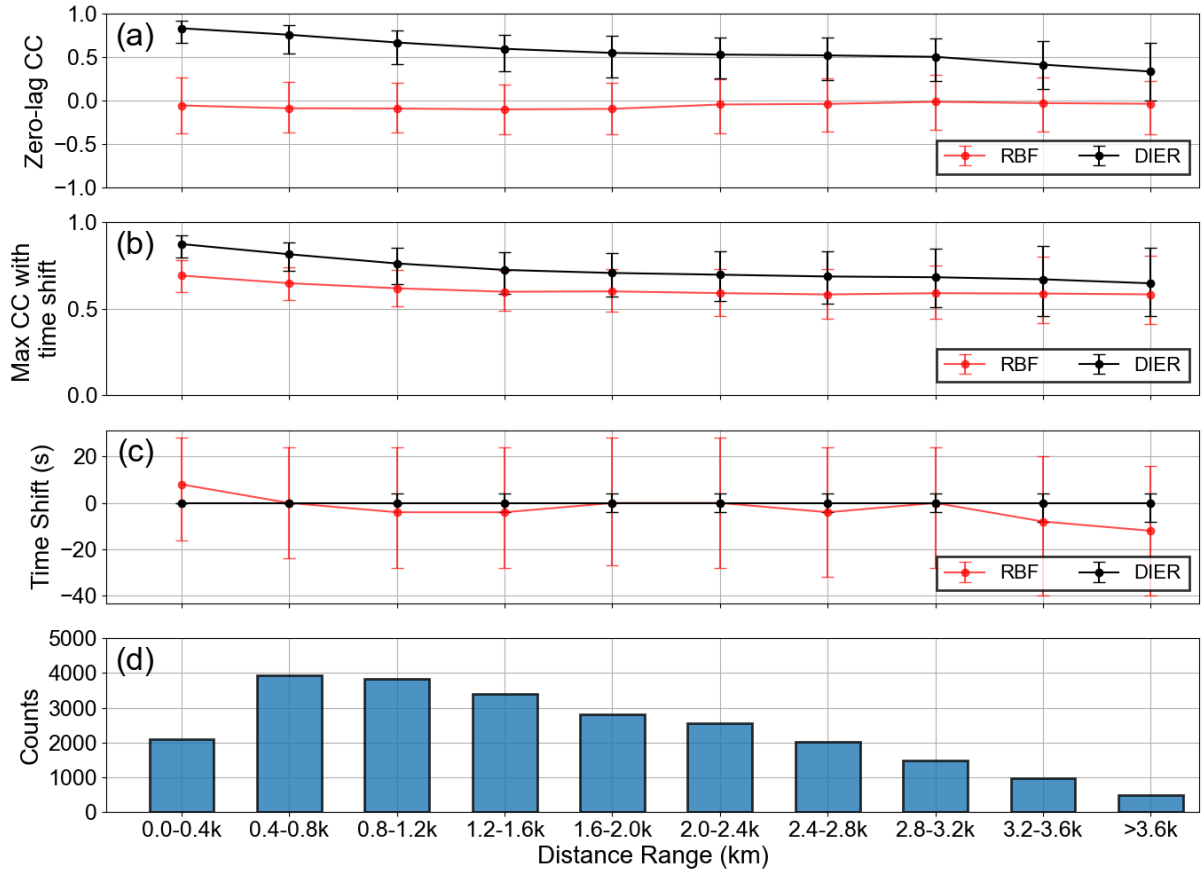

**Figure 6.** Quantitative comparison of interpolation performance between DIER and RBF-based approach for all EGFs in the test dataset. (a) Zero-lag cross-correlation coefficients between real EGFs and interpolated EGFs with RBF (red line) and DIER (black line) for varied distance ranges; the error bars indicate the 25th–75th percentile; (b) maximum cross-correlation coefficients with time shift between the real EGFs and interpolated EGFs; (c) corresponding time shifts for the maximum cross-correlation coefficients; (d) Histogram showing the number of EGFs in each distance range.

The delay or advance in interpolated EGFs can translate into errors in phase velocities at different periods. The discrepancies in phase at different periods between real and interpolated EGFs can be converted into time shifts by:

$$\delta t(T) = \frac{T}{2\pi} \delta\phi(T), \quad (16)$$

where $\delta\phi(T)$ is the phase difference at period $T$, and $\delta t(T)$ is the corresponding time shift. For a constant reference velocity $v_0 = 4 \text{ km s}^{-1}$, the phase-velocity perturbation ($\delta v(T)$) can be derived by:

$$\delta v(T) = v'(T) - v_0 = \frac{d}{d/v_0 + \delta t(T)} - v_0, \quad (17)$$

where $v'(T)$ is the estimated phase velocity from the spectrograms of interpolated EGFs (e.g., Fig. 5), and $d$ is the interstation distance. Fig. 7 illustrates the estimated phase-velocity perturbations for all EGFs in the test dataset. The interpolation results based on RBF (Fig. 7a) exhibit significantly scattered distributions, with deviations frequently exceeding ±0.15 km s$^{-1}$. Such large scattering of interpolation results indicates significant phase shifts, consistent with the large time shifts shown in Fig. 6(c). In comparison, the interpolation results by DIER exhibit compact and symmetric distributions around zero at different periods (Fig. 7b), with most deviations within ±0.05 km s$^{-1}$ of the reference velocity. Figs S2 and S3 show the phase-velocity perturbations for reference velocities $v_0$ of 3 km s$^{-1}$ and 5 km s$^{-1}$, respectively. Though increasing the reference velocity leads to larger perturbations, the contrast between the EGFs interpolated by DIER and RBF remains essentially unchanged.

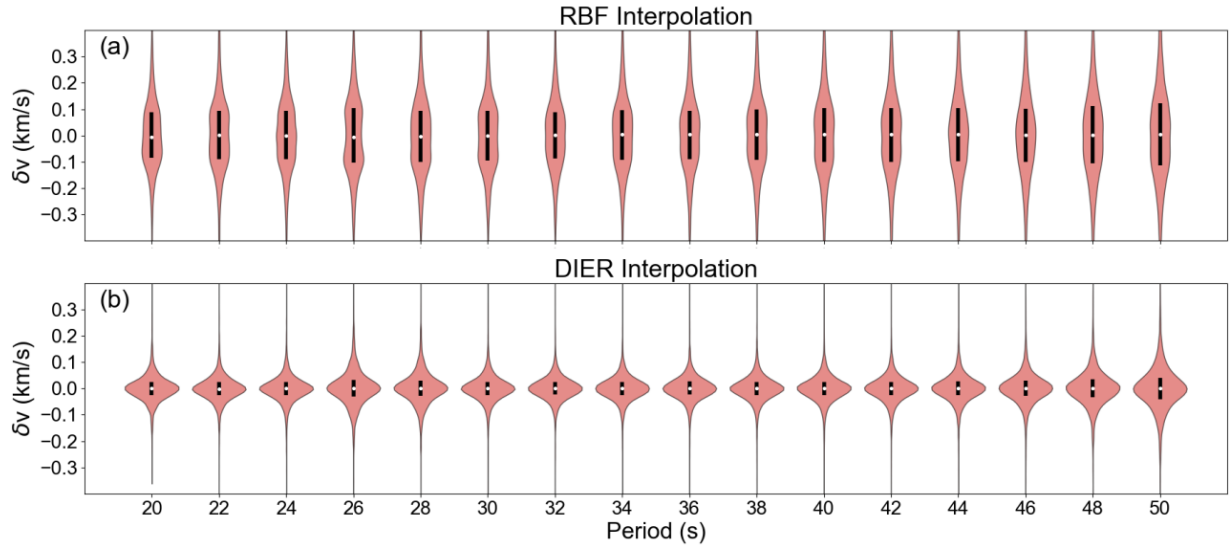

**Figure 7.** Violin plots for the distributions of phase-velocity perturbations at different periods for the EGFs interpolated with DIER and RBF. (a) Distribution of phase-velocity perturbations for EGFs interpolated by RBF, and (b) EGFs interpolated by DIER at different periods. The white dots indicate the median values, and the black bars represent the 25th–75th percentile ranges at each period.

## 4.2 Higher-resolution Phase Velocity Maps using Interpolated EGFs

Using the interpolated EGFs by DIER, we derive enhanced phase velocity maps and compared them with those derived from an actual denser array in the continental US. The actual stations in the training and testing datasets of DIER are denoted by the cyan dots in Fig. 8. For EGF interpolation, 1,046 virtual stations (red dots in Fig. 8) are uniformly placed across the continental U.S. at 0.9°×0.9° interval, and DIER is applied to reconstruct EGFs in the 20–50 s period range between each virtual station pairs. Note the frequency range is determined by the training dataset as discussed in Section 3. To guarantee that the interpolated EGFs contain sufficient dispersion information, we only generate EGFs for station pairs with interstation distances exceeding the longest wavelength of interest (Luo et al., 2015). Specifically, considering surface waves with a period of 50 s propagating at approximately 4 km s$^{-1}$, the minimum interstation distance for EGF interpolation is about 200 km.

Before deriving the phase velocity maps, we first assess the waveform similarity between the real and interpolated EGFs. Four real station pairs are randomly selected from various regions across the continental U.S. For each real pair, a nearby virtual station pair with an interpolated EGF is chosen for comparison (Fig. 9a). As shown in Fig. 9(b), the interpolated EGFs by DIER agree well with the real EGFs. Note the causal and anti-causal parts of the EGFs are asymmetrical in the real EGFs due to uneven distribution of noise sources (Nakata et al., 2019; Ni et al., 2022). Fig. 9(c) further compare the phase-velocity spectrograms for the real and interpolated EGFs in Fig. 9(b). Since the real and virtual station pairs are not entirely co-located, minor discrepancies are inevitable. Nevertheless, the strong overall similarity in the dispersive characteristics indicates that the interpolated EGFs at the virtual stations are suitable for constructing phase velocity maps.

To estimate surface-wave velocities at different periods, the interpolated EGFs are converted into phase-velocity spectrograms using the EGFAnalysisTimeFreq package (Yao et al., 2011; Zhao et al., 2023). In consideration of the large number of synthesized EGFs, we use DisperPicker (Yang et al., 2022) to automatically extract dispersion curves from the spectrograms. The effectiveness of DisperPicker has been demonstrated in real-time monitoring of a volcanic plumbing system based on ambient noise tomography (Stumpp et al., 2025). Also, to ensure the reliability of extracted dispersion curves, we apply a quality control procedure to the automatically picked curves, including picking confidence and curve length thresholding, constraints of curve continuity, and removal of outliers (Yang et al., 2022).

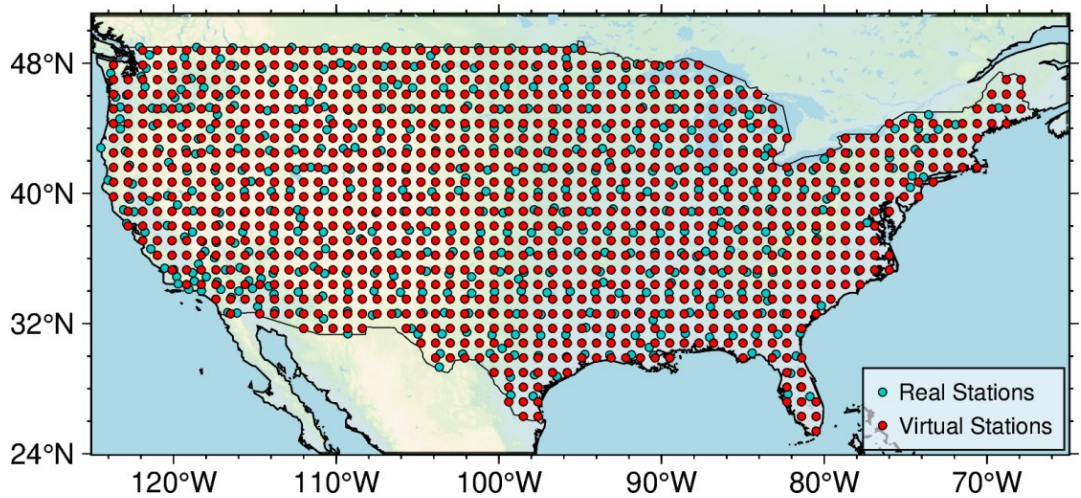

**Figure 8.** Distribution of real and virtual stations for surface-wave phase velocity tomography. The cyan dots denote the real seismic stations, and the red dots denote the virtual stations distributed uniformly at 0.9°×0.9°.

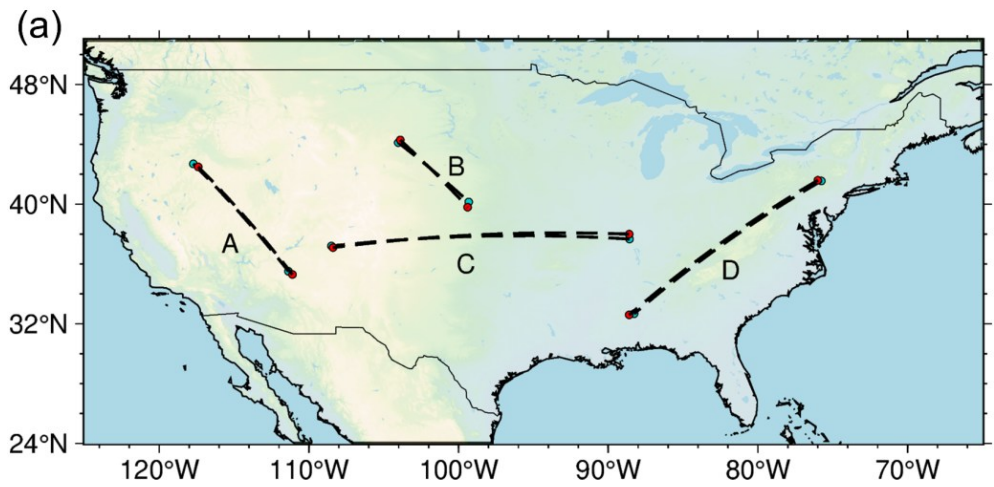
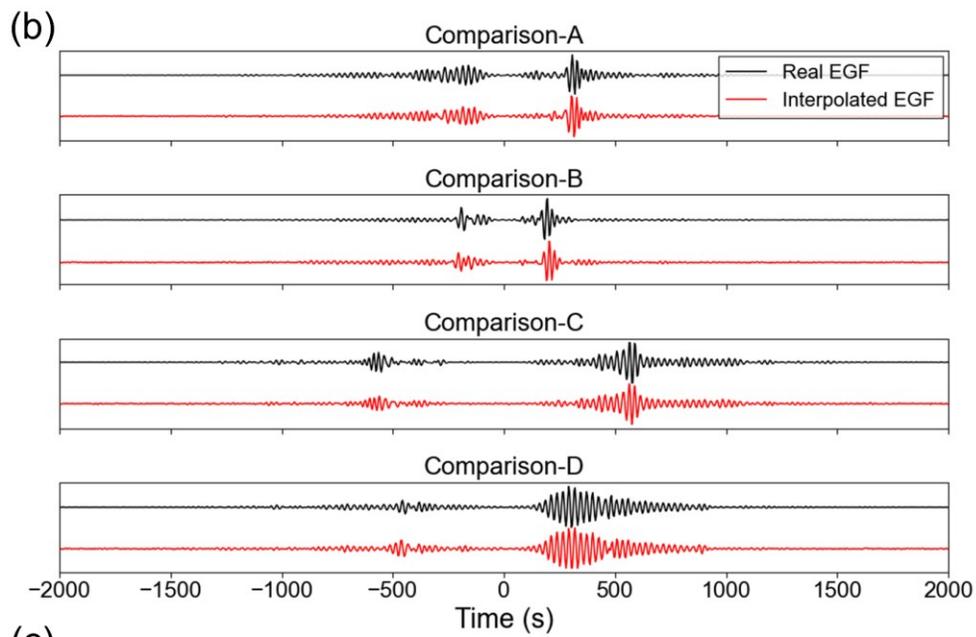
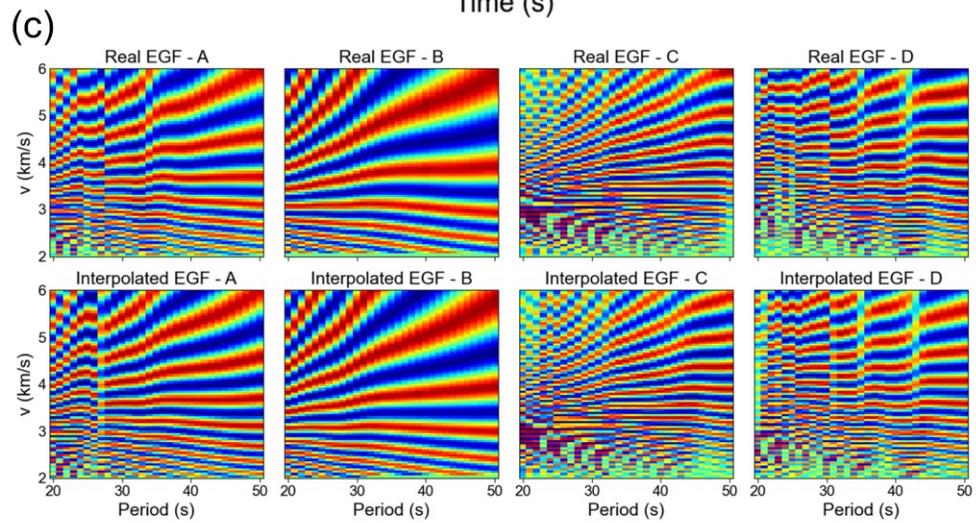

**Figure 9.** Representative examples illustrating the agreement between real and interpolated EGFs. (a) Distribution of four selected real and virtual station pairs in proximity. The cyan and red dots denote real and virtual stations, respectively, and the black dashed lines show the interstation surface-wave travel paths. (b) Waveform comparison for each example. The black traces represent the real EGFs, and the red traces represent the interpolated EGFs. (c) Comparisons of phase velocity spectrograms. The top row shows the spectrograms derived from the real EGFs, and the bottom row shows the spectrograms derived from the interpolated EGFs.

We use fast-marching surface tomography (FMST) (Rawlinson, 2005) to invert the automatically extracted phase velocity dispersion curves for phase velocity maps. We evaluate the phase velocity maps derived from DIER-interpolated EGFs against the 25-s and 40-s maps from the high-resolution, well-constrained continental U.S. phase velocity model (Babikoff & Dalton, 2019), which is based on dispersion data from 752 teleseismic events recorded by 1,831 USArray stations. Fig. 10(a) shows the phase velocity maps at 25 s derived from the 460 sparse stations; Fig. 10(b) shows the phase velocity map derived from interpolated EGFs with DIER, which exhibit a higher similarity with the reference map (Fig. 10c) than those derived directly from sparser observations (Fig. 10a). Figs (d)-(f) are similar to Figs (a)-(c), but for the 40-s phase velocity map. The discrepancy among different phase velocity maps can be distinctly revealed in three representative regions (R1-R3). Region R1 spans the northwestern states of the continental U.S., where the 25-s phase velocity map derived from sparse EGFs (Fig. 10a) exhibits a spotted pattern.

In comparison, R1 in the 25-s phase velocity map derived from DIER-interpolated EGFs (Fig. 10b) exhibits a continuous E-W trending low-velocity belt resembling the feature in the reference phase velocity map (Fig 10c). Region R2 spans the eastern states where the coverage of the actual stations is particularly sparse (cyan dots in Fig. 8). In Region R2, the NE-SW oriented high-velocity belt along the southeastern coast that is prominent in the reference model (Figs 10c and 10f) is largely absent in both the 25-s and 40-s phase velocity maps inverted from the sparse EGFs (Figs 10a and 10d). In comparison, the phase velocity maps derived from DIER-interpolated EGFs (Figs 10b and 10e) partially recover the high velocity belt. Region R3 spans the central U.S., which is characterized by a NE-SW trending high-velocity anomaly particularly pronounced in the 40-s phase velocity map in the reference model (Fig. 10f). The distinct high-velocity anomaly in this region is largely reconstructed using the EGFs interpolated with DIER (Fig. 10e). In comparison, this feature is absent in the phase velocity map derived from the EGFs acquired by the sparse network (Fig. 10d).

To further quantify the discrepancy among different phase velocity maps, we compute the structural similarity index measure (SSIM; Wang et al., 2004) between an inverted map and the corresponding map from the reference model (Table 1). The definition of SSIM is summarized in Appendix C. The 25-s and 40-s phase velocity maps derived from DIER-interpolated EGFs consistently achieve higher SSIM values than those obtained from the original sparse observations. The quantitative comparison confirms that the data-domain interpolation with DIER preserves accurate dispersion information and demonstrably improves the quality of the reconstructed phase velocity maps.

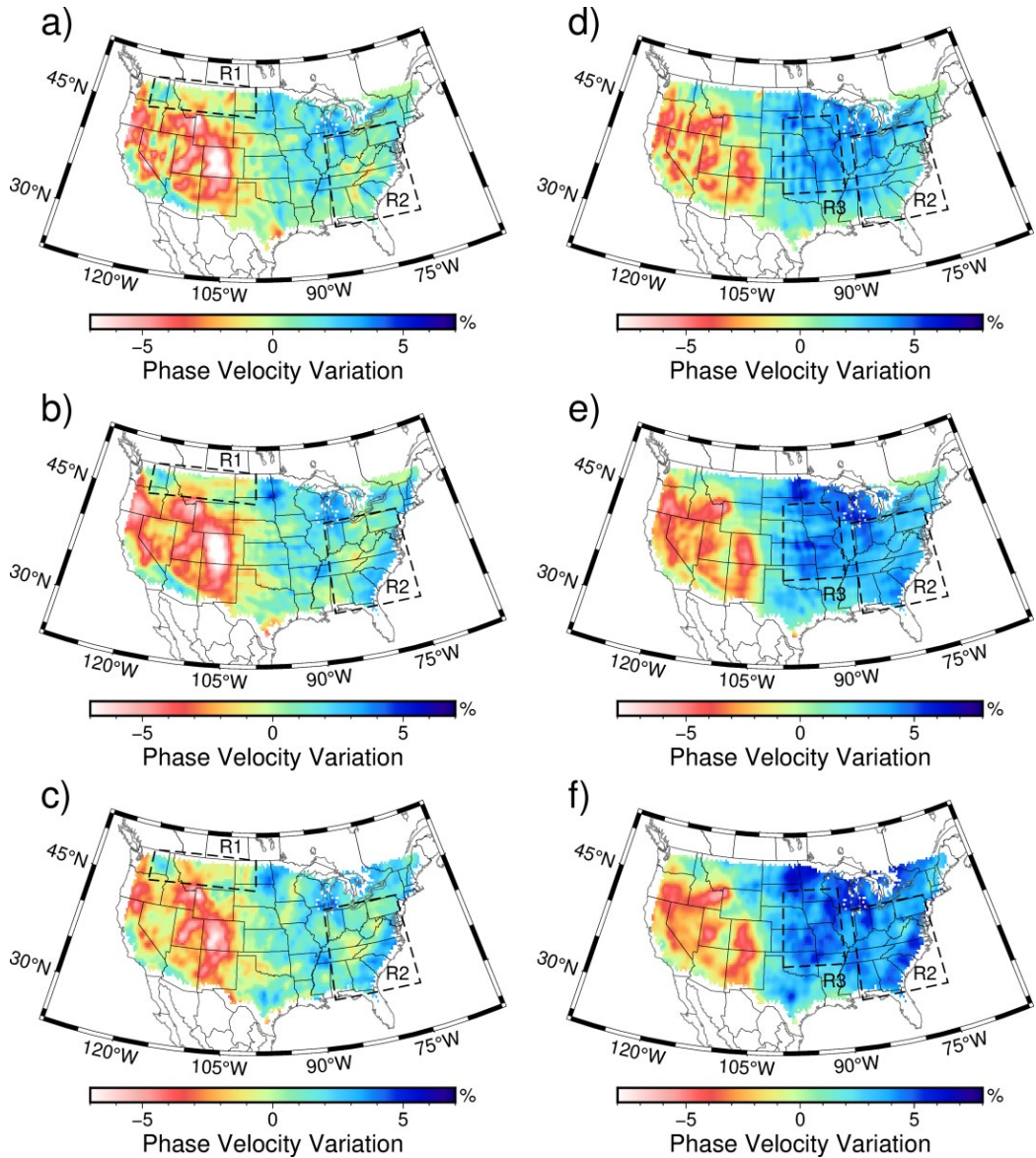

**Figure 10.** Comparisons of phase velocity maps for the continental U.S. (a) Phase velocity map at 25 s derived from EGFs collected by the sparse network. (b) presents the phase velocity map derived from EGFs interpolated with DIER at virtual stations as shown in Fig. 8. (c) presents the reference phase velocity map derived from an actual denser network (Babikoff & Dalton, 2019). (d)-(f) are similar to (a)-(c), but for phase velocity maps at 40 s. Black dashed boxes (R1-R3) mark

the regions discussed in the text.

**Table 1.** Structural similarity index measure (SSIM) values for the phase velocity maps derived from the EGFs collected by the sparse network and EGFs interpolated with DIER compared to the phase velocity maps from an actual denser network (Babikoff & Dalton, 2019).

| EGF origin | Period (s) | SSIM |
|---|---|---|
| Sparse network | 25 | 0.6485 |
|  | 40 | 0.6742 |
| Interpolation with DIER | 25 | 0.6851 |
|  | 40 | 0.7591 |

## 5. DISCUSSION

### 5.1 Choice of Coordinate Embedding Strategy

As mentioned in Section 2.2, coordinate embedding strategies can influence the generalization of INR-based models (e.g., Tancik et al., 2020; Müller et al., 2022). In this section, we compare the generalization capabilities of three different coordinate embedding strategies. One strategy is to directly feed the absolute latitude and longitude coordinates of interpolated station pairs into the MLP layer. Alternatively, relative positions between station pairs can be embedded, and the coordinate difference is used as the conditional guidance, i.e., $f_{\theta_c}(\mathcal{C}) = f_{\theta_c}(\phi_i, \lambda_i) - f_{\theta_c}(\phi_j, \lambda_j)$. A third strategy adopts the Fourier feature mapping (Tancik et al., 2020) to project the normalized coordinates into a high-dimensional vector, which can assist the network to capture high-frequency details in the reconstructed data. In this strategy, station coordinates $C = \{\phi_i, \lambda_i, \phi_j, \lambda_j\}$ are first normalized to interval [0,1] using min-max normalization when performing Fourier feature mapping:

$$\bar{C} = \frac{C - C^{min}}{C^{max} - C^{min}}, \quad (18)$$

where $\bar{C}$ is the normalized coordinates, $C^{min}$ and $C^{max}$ denote the minimum and maximum values of each coordinate dimension. The normalization can mitigate scale imbalance between latitude and longitude values, and to ensure that the subsequent sinusoidal projections receive inputs within a consistent numerical range. The normalized coordinates $\bar{C} = \{\bar{\phi}_i, \bar{\lambda}_i, \bar{\phi}_j, \bar{\lambda}_j\}$ are then mapped into a high-dimensional Fourier basis $C'$:

$$C' = [\cos(w_K x), \sin(w_K x)]_{x \in \bar{C}, k=1,\dots,K}, \quad (19)$$

where $K$ denotes the total number of frequency components. We set $K = 40$ and $w_i = 1.25^i \times \pi$ following Gao *et al.* (2025). The frequency-encoded features $C'$ are then processed by the MLP $f_{\theta_C}(C')$ before input into the U-Net.

Fig. 11 shows the loss curves for the training and testing datasets of three different embedding strategies. For the training dataset, Fourier feature mapping (Eq. 19) outperforms both absolute and relative position embedding strategies, a result consistent with previous findings that Fourier encodings, compared to direct feeding of raw coordinates, can markedly enhance the ability of a neural network to represent high-frequency variations in high-dimensional seismic data (Gao et al., 2025). However, the proposed DIER framework leverages diffusion models for data reconstruction, which exhibit significantly greater representational capacity for complex waveform details and can accurately fit the training datasets even when using simple absolute coordinates. In such cases, incorporating Fourier feature mapping tends to promote overfitting, resulting in degraded generalization performance on the test dataset. In other words, the additional high-frequency data fitting ability conferred by Fourier feature mapping may be unnecessary, and could even impair overall generalization performance. In contrast, absolute position embeddings consistently yield the most robust interpolation results with the lowest test errors.

Fig. 12 shows representative EGFs reconstructed by the three embedding strategies. The reconstructed EGFs with the Fourier feature mapping or relative position embedding tend to deviate from the real EGFs, whereas the absolute position embedding yields the closest match with accurate phase and amplitude characteristics. These illustrative examples confirm that absolute position embedding provides the most reliable generalization capability.

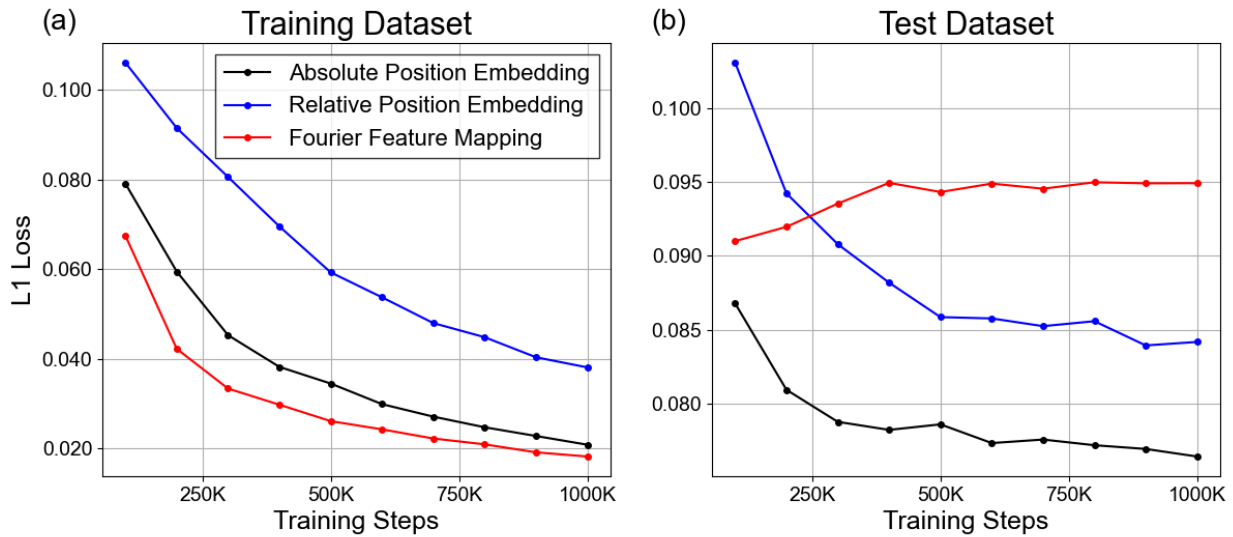

**Figure 11.** Comparison of different coordinate embedding strategies for DIER. (a) Loss curves for the training dataset, and (b) loss curves for the test dataset in L1-norm over one million training

steps. The black line indicates the loss curve for absolute position embedding, the blue line indicates the loss curve for the relative position embedding, and the red line indicates the loss curve for the Fourier feature mapping.

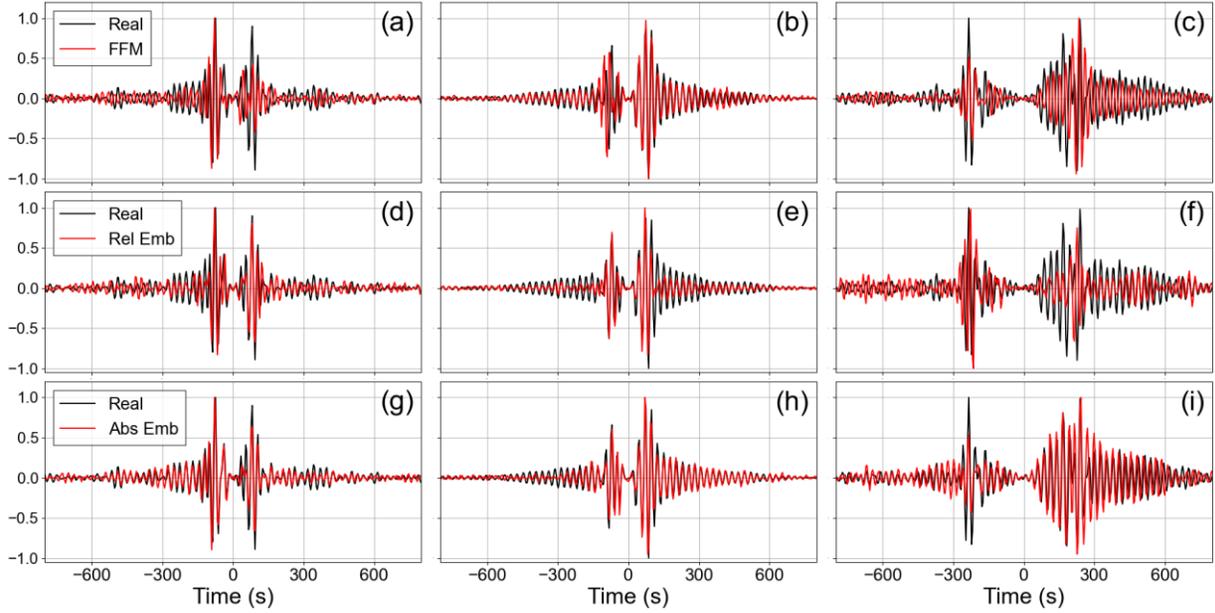

**Figure 12.** EGF interpolation results using different coordinate embedding strategies. Each row compares the real EGFs (black) with interpolated EGFs (red) using a specific coordinate embedding strategy at different station pairs. (a-c) Comparison of real EGFs with EGFs interpolated with Fourier feature mapping, (d-f) with EGFs interpolated with relative position embedding, and (g-i) with EGFs interpolated with absolute position embedding. Each column corresponds to a different station pair.

## 5.2 Limitations and future work

The above discussion highlights the critical influence of conditional guidance in the INR framework. Excessive coordinate encodings may introduce unnecessary inductive bias (Yang et al., 2023) and counteract generalization capability of the neural network. Future studies could investigate other coordinate embedding strategies (Müller et al., 2022) or integrate physics-informed priors (Karniadakis et al., 2021) to guide the EGF generation process with increased fidelity.

As shown in Fig. 10, while the large-scale structures in the phase velocity maps reconstructed from DIER-interpolated EGFs agree well with those in the reference maps, the fine-scale details are noticeably smoother than those derived from real EGFs recorded by a denser array (Babikoff & Dalton, 2019). The smoothing behavior in the model domain suggests an implicit regularization effect inherent to the data-domain denoising-based interpolation process. Future work could focus on designing adaptive regularization strategies capable of preserving localized structural anomalies during data interpolation.

In addition, the proposed method uses DDPM as the backbone for implicit representation, which aims to accurately model the dispersive surface-wave propagation in EGFs. Unlike the large-scale pretraining paradigms based on massive data, EGFs are typically limited in quantity and exhibit distinct in-domain heterogeneity (Wang et al., 2025) in varied regions. In such cases, diffusion models can benefit from repeatedly learning the limited data samples than other data-driven generative approaches like autoregressive models (Prabhudesai et al., 2025). Future efforts could focus on optimizing the diffusion process to further improve interpolation quality by adopting more expressive network backbones such as attention-based U-Nets or diffusion transformers (Rombach et al., 2022; Peebles & Xie, 2023). Moreover, as generative models, DDPMs are inherently stochastic and could produce varied waveforms (Chen et al., 2024). In the current work, we perform a single sampling for each virtual station pair using DIER. Repeated sampling at the same coordinates can further reduce the misfit between real and interpolated EGFs (Fig. 13). However, this improvement comes at a significant computational cost, since each sampling requires multiple denoising steps. Adopting more efficient denoising schedules (e.g., Song et al., 2020) would be valuable for rapid estimation of interpolation uncertainty.

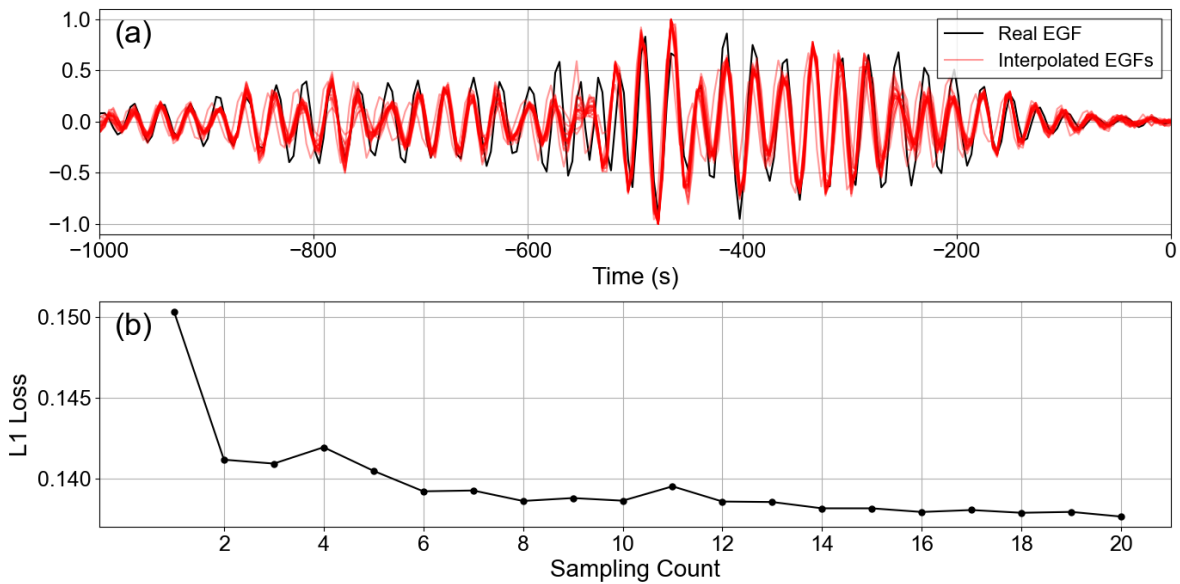

**Figure 13.** Example of repeated sampling at a station pair using DIER. (a) Real EGF (black) and 20 EGFs interpolated by DIER (red). (b) L1 loss of waveform discrepancy in (a) as a function of the total sampling count. For each data point, the loss is computed between the real EGF and the median of all generated EGFs up to a particular sampling count.

# 6. CONCLUSION

In this study, we propose a novel framework named diffusion-assisted implicit EGF representation (DIER), which integrates advanced generative modeling with implicit neural representation for accurate interpolation of dispersive ambient noise empirical Green's functions. By conditioning the diffusion process on station coordinates, DIER allows self-supervised reconstruction of irregularly sampled five-dimensional EGF wavefields without requiring labeled or synthetic training datasets. Compared with the conventional radial basis function-based interpolation, EGFs interpolated with DIER exhibit superior waveform fidelity, more accurate phase information and dispersion characteristics. In addition, the phase-velocity tomography based on DIER-interpolated EGFs from a sparse network closely matches a reference model obtained from a much denser network. The proposed approach for dispersive data interpolation is promising for improving stability and resolution of surface wave tomography in regions with limited observations. Beyond EGF interpolation, the proposed DIER framework offers a general paradigm for applying diffusion-based implicit representations to other types of irregularly sampled, high-dimensional scientific data. We anticipate that the proposed framework will inspire new approaches for modeling and understanding complex high-dimensional physical processes in seismology and across the broader geosciences.

# ACKNOWLEDGEMENTS

This research was supported by National Natural Science Foundation of China (Grant Nos. U2139204, 42488301), and by the National Key Research and Development Project of China (Grant No. 2022YFF0800701). GPUs used in this study for training the neural networks are provided by the Hefei Advanced Computing Center on the HG architecture. The authors thank Prof. Haijiang Zhang for his helpful suggestions in conceptualizing this paper. They also thank Dr. Peng Zou for his suggestions regarding the RBF-based interpolation.

# DATA AVAILABILITY STATEMENT

EGF data used in this study can be downloaded at the Incorporated Research Institutions for Seismology (IRIS) data center (https://ds.iris.edu/ds/products/globalempiricalgreenstensors/). The source code of DIER can be accessed at https://github.com/Billy-Chen0327/DIER.

# SUPPORTING INFORMATION

**Figure S1.** Comparison of dispersion characteristics of observed and interpolated EGFs for the other representative example with inferior reconstruction quality by RBF and DIER. The rest is same as Fig. 5.

**Figure S2.** Violin plots for the distributions of phase-velocity perturbations at different periods for the EGFs interpolated with DIER and RBF, assuming a reference velocity of 3 km s$^{-1}$. The rest is same as Fig. 7.

**Figure S3.** Violin plots for the distributions of phase-velocity perturbations at different periods for the EGFs interpolated with DIER and RBF, assuming a reference velocity of 5 km s$^{-1}$. The rest is same as Fig. 7.

Wu, L., Li, J., Bao, S., & Gong, Q. (2024). Visualization analysis of ambient seismic noise research. *Frontiers in Earth Science*, *12*, 1452324.

Xu, J., Li, H., & Zhou, S. (2015). An overview of deep generative models. *IETE Technical Review*, *32*(2), 131-139.

Yang, S., Zhang, H., Gu, N., Gao, J., Xu, J., Jin, J., ... & Yao, H. (2022). Automatically extracting surface-wave group and phase velocity dispersion curves from dispersion spectrograms using a convolutional neural network. *Seismological Research Letters*, *93*(3), 1549-1563.

Yang, J., Pavone, M., & Wang, Y. (2023). Freenerf: Improving few-shot neural rendering with free frequency regularization. In *Proceedings of the IEEE/CVF conference on computer vision and pattern recognition* (pp. 8254-8263).

Yao, H., van Der Hilst, R. D., & De Hoop, M. V. (2006). Surface-wave array tomography in SE Tibet from ambient seismic noise and two-station analysis—I. Phase velocity maps. *Geophysical Journal International*, *166*(2), 732-744.

Yao, H., Gouedard, P., Collins, J. A., McGuire, J. J., & van der Hilst, R. D. (2011). Structure of young East Pacific Rise lithosphere from ambient noise correlation analysis of fundamental- and higher-mode Scholte-Rayleigh waves. *Comptes Rendus. Géoscience*, *343*(8-9), 571-583.

Yeeh, Z., Song, Y., Byun, J., Seol, S. J., & Kim, K. Y. (2020). Regularization of multidimensional sparse seismic data using Delaunay tessellation. *Journal of Applied Geophysics*, *174*, 103877.

Yuan, P., Wang, S., Hu, W., Nadukandi, P., Botero, G. O., Wu, X., ... & Chen, J. (2022). Self-supervised learning for efficient antialiasing seismic data interpolation. *IEEE Transactions*
47

# APPENDIX A

## Interpolation based on radial basis functions

As an interpolation method for unmeshed data, radial basis function-based (RBF) interpolation (Fasshauer, 2007) is adopted as the baseline in our analysis. For rigorous comparison with DIER, we perform RBF-based interpolation at each time sample of the EGFs. $C_n \in \mathbb{R}^4$ ($n = 1, \ldots, N$) represents the station coordinates for $N$ observational station pairs:

$$C_n = \{\phi_n, \lambda_n, \phi'_n, \lambda'_n\}, \quad (A1)$$

where $(\phi_n, \lambda_n)$ and $(\phi'_n, \lambda'_n)$ are the latitudes and longitudes of the two stations in the $n^{th}$ pair. For the coordinates $C$ of an arbitrary station pair, we use $\hat{G}(C, \tau)$ to represent the EGF at time sample $\tau$, and the RBF-based interpolant can be written as:

$$\hat{G}(C, \tau) = \sum_{n=1}^{N} a_n(\tau) \mathcal{R}(\|C - C_n\|_2) + \sum_{m=1}^{M} b_m(\tau) p_m(C), \quad (A2)$$

where $\mathcal{R}$ is the radial basis function, and $p_m$ is a monomial that spans the space of polynomials with the order $m$; for each time sample $\tau$, The coefficient vectors $\boldsymbol{a} = [a_1(\tau), \ldots, a_N(\tau)]$ and $\boldsymbol{b} = [b_1(\tau), \ldots, b_M(\tau)]$ are obtained by solving the following linear equations:

$$(K + \sigma^2 I)a + Pb = d, \quad (A3)$$

and

$$P^T a = 0, \quad (A4)$$

where $\sigma$ is a smoothing parameter, $I$ is the identity matrix, $K$ is a matrix with entry $K_{mn} = \mathcal{R}(\|C_m - C_n\|_2)$, $P$ is a matrix with entry $P_{nm} = p_m(C_n)$, $d = [\hat{G}(C_1, \tau), \ldots, \hat{G}(C_N, \tau)]$ collects the real EGF at time sample $\tau$ for different station pairs. The hyperparameter settings in the RBF-

based interpolation follow Zou et al. (2024), in which the smoothing parameter $\sigma$ is fixed to 0.2, and the order of polynomials $M$ is set to 1. The second-order polyharmonic spline function is adopted as the radial basis function $\mathcal{R}$:

$$\mathcal{R}(r) = (\varepsilon r)^2 \log \varepsilon r, \quad (A5)$$

where $\varepsilon$ is the shape parameter determining how rapidly the basis function decays with distance $r$. We set $\varepsilon = 50$ in this study.

# APPENDIX B

## Definitions of Correlation-Based Similarity Metrics

In this study, three correlation-based metrics are used to quantify the similarity between interpolated and real EGFs: the zero-lag cross-correlation coefficient, the maximum cross-correlation coefficient, and the corresponding time shift. Let $\hat{G}(t)$ and $\hat{G}'(t)$ denote the real and interpolated EGF, respectively, the time-lag cross-correlation coefficient is defined as:

$$\rho(\tau) = \frac{\int_{-\infty}^{\infty} \left(\hat{G}(t) - \overline{\hat{G}(t)}\right)\left(\hat{G}'(t+\tau) - \overline{\hat{G}'(t)}\right) dt}{\sqrt{\int_{-\infty}^{\infty} \left(\hat{G}(t) - \overline{\hat{G}(t)}\right)^2 dt \int_{-\infty}^{\infty} \left(\hat{G}'(t) - \overline{\hat{G}'(t)}\right)^2 dt}}, \quad （B1）$$

where $\overline{\hat{G}(t)}$ and $\overline{\hat{G}'(t)}$ are the temporal means of the real and interpolated EGFs, respectively. The maximum cross-correlation coefficient $\rho^{max}$ is defined as the maximum value of $\rho(\tau)$ over all possible time lags $\tau$:

$$\rho^{max} = \max_{\tau} \rho(\tau), \quad （B2）$$

and the corresponding time shift $\tau^*$ is the time lag at which $\rho^{max}$ is attained:

$$\tau^* = \operatorname*{argmax}_{\tau} \rho(\tau). \quad （B3）$$

In addition, we use the zero-lag cross-correlation coefficient to describe the correlation between the two waveforms without any time shift:

$$\rho(0) = \rho(\tau)|_{\tau=0}. \quad （B4）$$

# APPENDIX C

## Definitions of the structural similarity index

We adopt structural similarity index (SSIM) (Wang et al., 2004) to evaluate the similarity between an inverted phase velocity map and the reference phase velocity map. SSIM is defined as:

$$SSIM = \frac{(2\mu_1\mu_2 + C_1)(2\sigma_{12} + C_2)}{(\mu_1^2 + \mu_2^2 + C_1)(\sigma_1^2 + \sigma_2^2 + C_2)}, \quad (C1)$$

where $\mu_1$ and $\sigma_1$ are the mean and standard deviation of the inverted phase velocity map, $\mu_2$ and $\sigma_2$ are the mean and standard deviation of the reference velocity map, the term $\sigma_{12}$ denotes the covariance between inverted and reference phase velocities, $C_1$ and $C_2$ are small positive constants for stabilizing the computation when $(\mu_1^2 + \mu_2^2)$ and $(\sigma_1^2 + \sigma_2^2)$ are close to zero. Following van der Walt et al. (2014), $C_1$ and $C_2$ are set to $(0.01*L)^2$ and $(0.03*L)^2$, respectively, where $L$ denotes the data range, i.e., the difference between the minimum and maximum values of the reference velocity model.

Supplement to

**Accurate Interpolation of Ambient Noise Empirical Green's Functions by Denoising**

**Diffusion Probabilistic Model and Implicit Neural Representation**

By Guoyi Chen, Junlun Li* and Bao Deng

**Supplemental Figures**

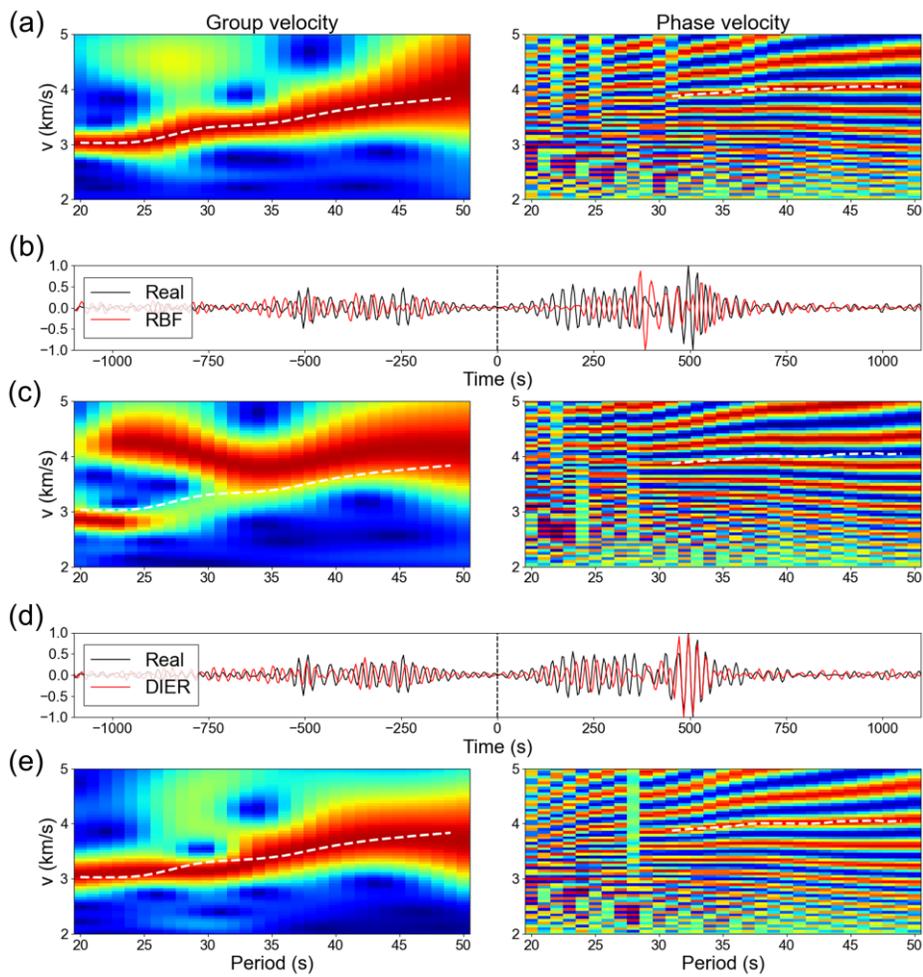

**Figure S1.** Comparison of dispersion characteristics of observed and interpolated EGFs for the other representative example with inferior reconstruction quality by RBF and DIER. The rest is same as Fig. 5.

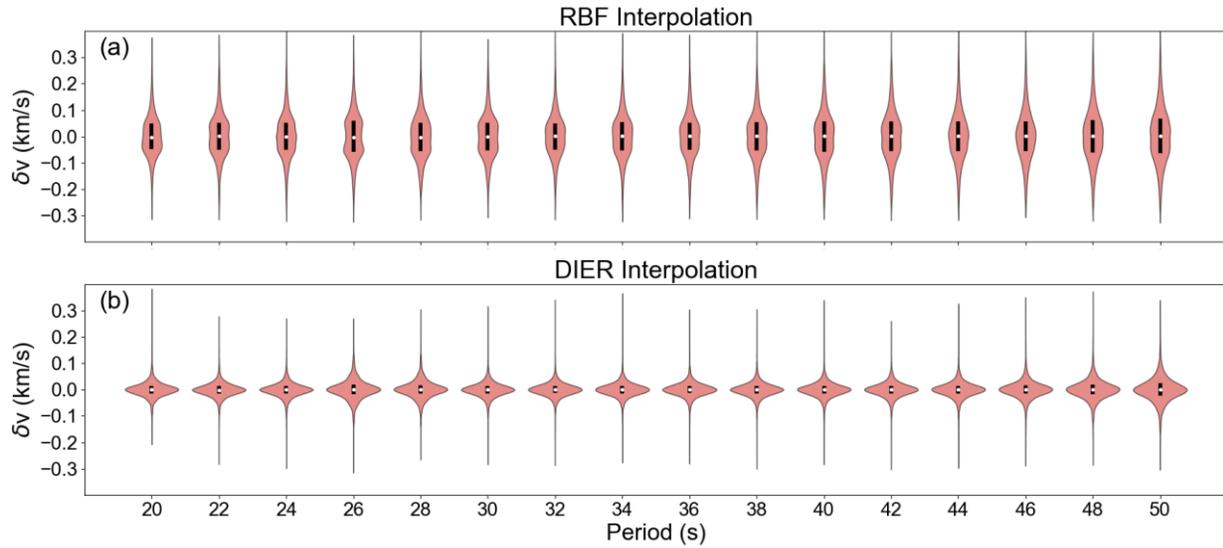

**Figure S2.** Violin plots for the distributions of phase-velocity perturbations at different periods for the EGFs interpolated with DIER and RBF, assuming a reference velocity of 3 km s$^{-1}$. The rest is same as Fig. 7.

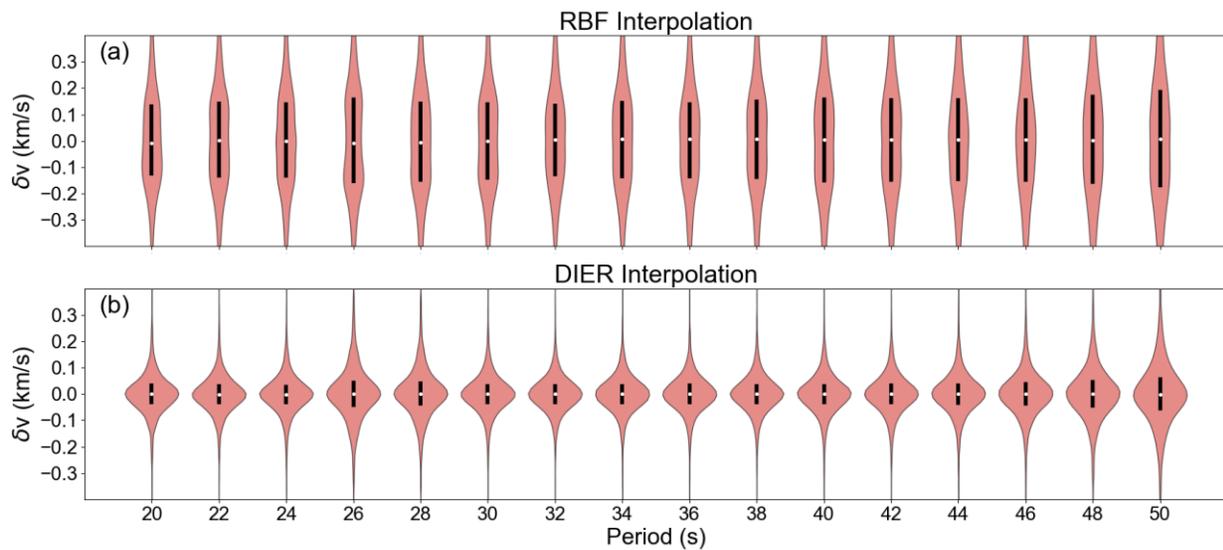

**Figure S3.** Violin plots for the distributions of phase-velocity perturbations at different periods for the EGFs interpolated with DIER and RBF, assuming a reference velocity of 5 km s$^{-1}$. The rest is same as Fig. 7.